\documentclass[%
 reprint,
unsortedaddress,
nofootinbib,
 amsmath,amssymb,
 aps,
floatfix,
]{revtex4-2}

\usepackage{comment}
\usepackage{graphicx}% Include figure files
\usepackage{dcolumn}% Align table columns on decimal point
\usepackage{bm}% bold math
\usepackage[dvipsnames]{xcolor}
\begin{document}

\preprint{APS/123-QED}

\title{Multi-dimensional memory in sheared granular materials}

\author{Chloe W. Lindeman}
\email{cwlindeman@uchicago.edu}
\affiliation{%
 Department of Physics and The James Franck and Enrico Fermi Institutes \\ 
 The University  of  Chicago, 
 Chicago,  IL  60637,  USA.\\
\\
}%

\date{\today}% It is always \today, today,
 
\begin{abstract}
%The ability of disordered materials like jammed packings to form a memory of previous deformation has received a great deal of attention in recent years. However, relatively little work has been done on the effect of training in a higher-dimensional phase space where multiple types of deformation play a role. 
To explore what features of multi-dimensional training can be remembered in granular materials, the response of a small, two-dimensional packing of hydrogel spheres to two independent types of shear is measured. Packings are trained via the application of several identical shear cycles, either of a single shear type or combinations of the two types. The memory is then read out using a standard protocol capable of revealing memories as a cusp at the point where readout reaches the training strain. The ability to read out a memory is sensitive not only to the type of deformation applied but also to the order in which different types of training are performed. These results underscore the importance of thinking of memories not as single remembered value (amplitude) but as a learned path through phase space.%, and they suggest a need for more general readout protocols.
\end{abstract}

\maketitle  

\section*{Introduction}

%\textit{Introduction} --- 

Despite their complex and disordered nature, granular materials are capable of remembering information about past deformations. On the one hand, their sensitivity to history is not surprising: any system with a plastic component to its response will by definition not return to its initial condition after a perturbation is removed. On the other hand, the lack of an ordered configuration to compare against might suggest difficulty reading out such a memory. Yet clear signatures have been demonstrated experimentally not only for the memory of a direction of driving~\cite{toiya2004transient, gadala1980shear} but also for the memory of a single shear amplitude~\cite{mukherji2019strength} and of multiple shear amplitudes~\cite{keim2020global}. It was also shown in simulations that the memory of a waveform can be recovered~\cite{candela2023complex}.

The origin of amplitude memory in particular is the subject of a great many studies~\cite{keim2019memory, paulsen2024mechanical}, ranging from comparisons of granular systems with toy models~\cite{fiocco2014encoding, mungan2019cyclic, keim2020global, keim2021multiperiodic, szulc2022cooperative, kumar2022mapping, movsheva2023granular} to direct molecular dynamics simulations~\cite{fiocco2014encoding, royer2015precisely, mungan2019networks, benson2021memory} to experiments across a variety of systems~\cite{mukherji2019strength, keim2020global, schwen2020embedding, benson2021memory, galloway2022relationships, chen2023microstructural}.
%The origin of amplitude memory in particular is the subject of a great many computational studies, both direct molecular dynamics simulations~\cite{fiocco2014encoding, royer2015precisely, mungan2019networks, benson2021memory} as well as comparisons of granular systems with toy models~\cite{fiocco2014encoding, keim2021multiperiodic, szulc2022cooperative, keim2020global, movsheva2023granular, mungan2019cyclic, kumar2022mapping}. Experimental realizations are limited to a handful of systems including bubble rafts, acrylic grains, and colloidal or nanocolloidal gels and glasses ~\cite{mukherji2019strength, keim2020global, schwen2020embedding, benson2021memory, chen2023microstructural}. 
It is generally understood that the formation of memory arises from plastic events (particle rearrangements) that occur as the material is strained. Strikingly, these plastic events become more repeatable over time, often reaching a periodic orbit after just a few cycles of driving. Clearly, this transition to a ``stroboscopic steady state,'' where particles return to the same positions at the end of every cycle, is closely linked to the formation of a memory of driving. 

Although a great deal of work has been done to identify~\cite{morse2020differences, lindeman2024minimal}, characterize~\cite{xu2017instabilities, xu2023discontinuous}, and predict~\cite{richard2020predicting, manning2011vibrational, cubuk2015identifying} these rearrangement events, there exist relatively few studies on the effect of the $\textit{type}$ of driving~\cite{gendelman2015shear,  patinet2016connecting, barbot2018local, schwen2020embedding, xu2021atomic, adhikari2023encoding}. Moreover, the experimental studies of this type of granular memory are limited to a handful of realizations. Here I explore the effect of driving an experimental two-dimensional granular material with both shear degrees of freedom: one in which the walls of the bounding box remain perpendicular but the relative lengths change, and one in which the lengths of all four walls remain equal to one another while the angles between the walls change. I adopt the notation of~\cite{scheibner2020odd} and refer to these as $S1$ and $S2$ shear, respectively. (In three dimensions there are five shear degrees of freedom, complicating the picture substantially.)

\begin{figure}
\centering
\includegraphics[width=6cm]{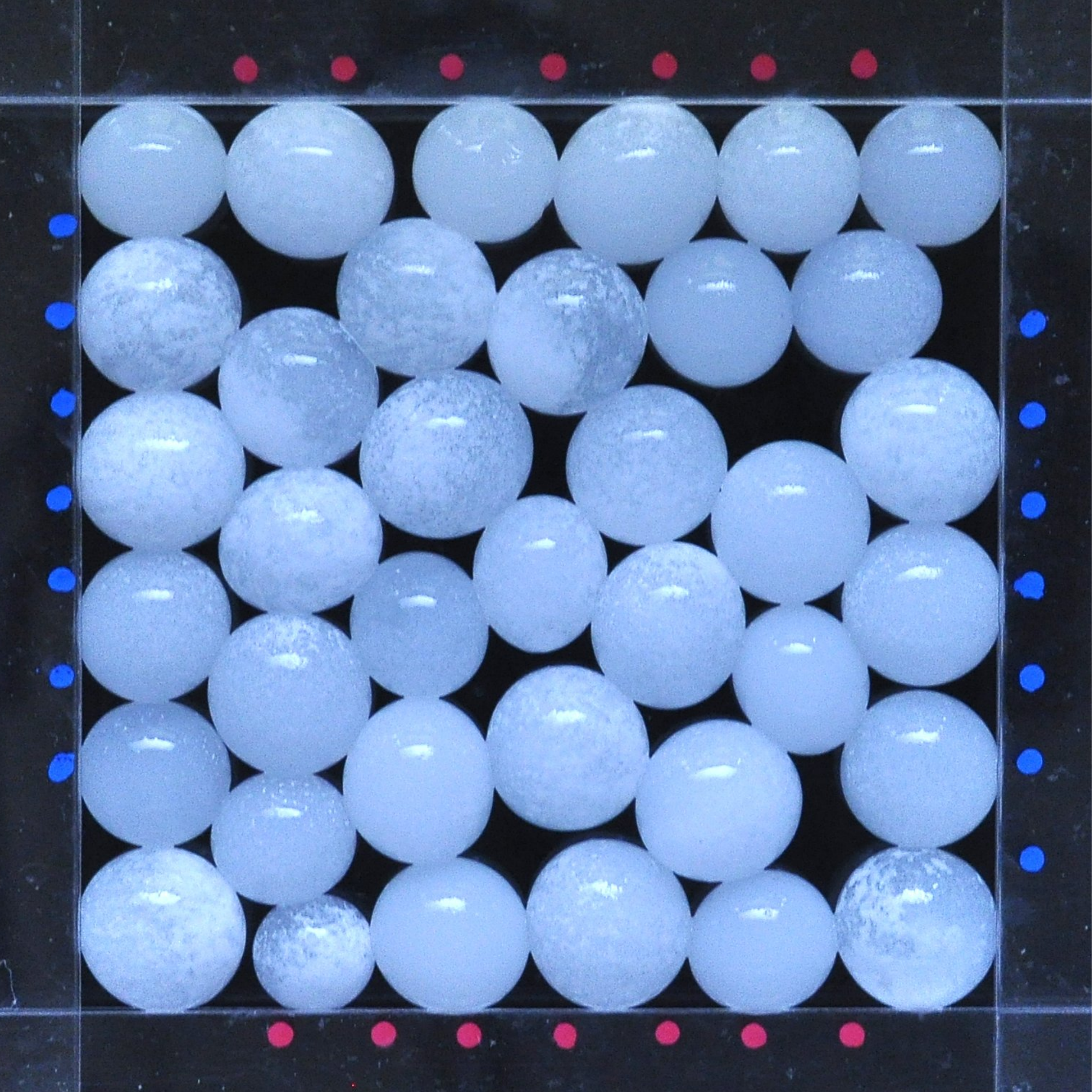}%
\caption{Two-dimensional packing of hydrogel spheres constrained by clear acrylic walls.
}
\label{fig:photo}
\end{figure}

\begin{figure}
\centering
\includegraphics[width=8.6cm]{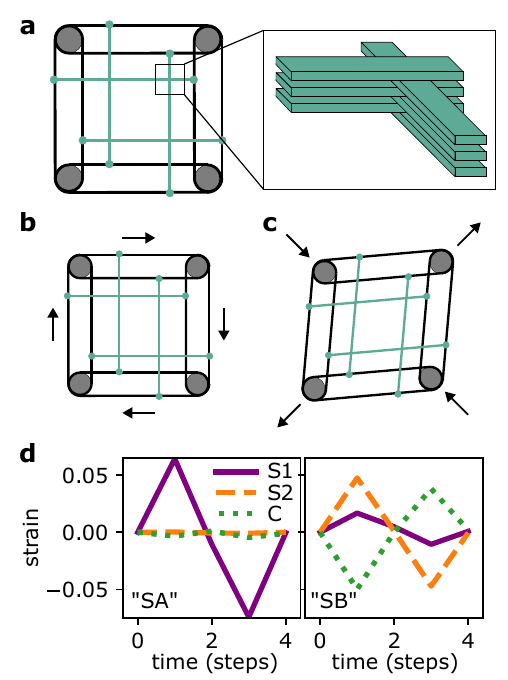}%
\caption{(a) Schematic of shear cell. Cell walls (green) are layered as shown on right, allowing intersecting walls to move through one another. Circles at the corners (grey) are pulleys and black ovals are belts; walls are secured to belts only where dots are shown. Particles are constrained to the inner square region. (b) Shear type $SA$ is applied by rotating the pulleys so that walls move in and out as shown. (c) Shear type $SB$ is applied by moving the whole pulley system as shown. Pulley axles are constrained to move along the diagonals. (d) Definition of shear types $SA$ and $SB$ in terms of $S1$, $S2$, and compression $C$. 
}
\label{fig:schematic}
\end{figure}

While the response of a granular system to multiple applied cycles of one type of shear --- the ability to remember not just one but two or more shear amplitudes --- is already surprising, the effect of training in a higher-dimensional phase space remains untested.
In this work, packings of roughly 35 hydrogel spheres like the one shown in Fig.~\ref{fig:photo} are trained with cycles of shear and the strength of memory is measured directly via a standard sequential readout protocol~\cite{adhikari2018memory, keim2020global, keim2022mechanical}. Training and readout are accomplished using the custom-built setup shown schematically in Fig.~\ref{fig:schematic}(a-c), which can apply two independent strain types $SA$ and $SB$ as defined in Fig.~\ref{fig:schematic}(d).
The results indicate that memories can be easily encoded in this system for one-dimensional driving but that the picture becomes more complicated when the full two-dimensional phase space is explored, revealing a dependence on the full history of past deformations.

\section*{Protocol}

A custom setup is designed to enable arbitrary application of $S1$ and $S2$. Application of $S1$, which requires the walls to change length, is accommodated by confining the particles with stacked acrylic slats, interleaved as shown in the inset to Fig.~\ref{fig:schematic}a so that the walls are capable of passing through one another at their corners. This allows for arbitrary box shape, not even providing the constraint that opposite walls remain parallel. The rest of the apparatus shown in Fig.~\ref{fig:schematic}a is therefore dedicated to constraining the walls to a parallelogram shape. 

\begin{figure}
\centering
\includegraphics[width=8.6cm]{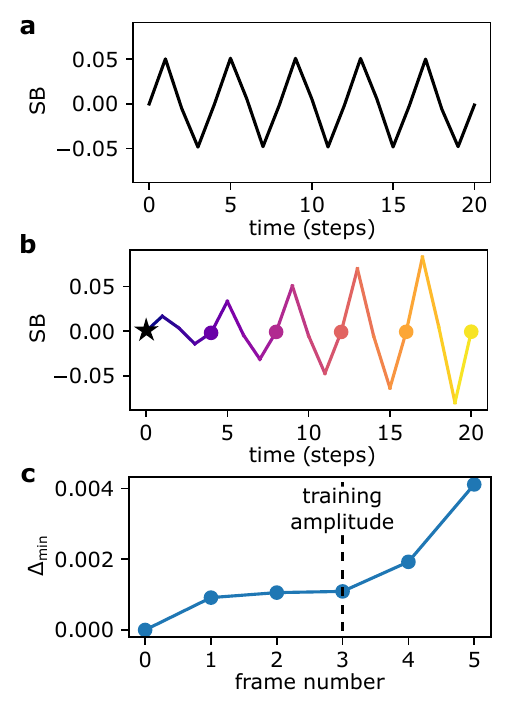}%
\caption{
(a) Training protocol: repeated shear cycles with a fixed shear type (here, $SB$) and amplitude.
(b) Readout protocol: repeated shear cycles beginning with very small amplitude and ramping up. Particle positions are measured at the end of each cycle (dots) and compared with the trained configuration (black star). Dark to light color transition shows progression of steps, redundant here but consistent with phase-space pictures in which time is not explicitly shown.
(c) Readout after the training shown in (b). Signal exhibits characteristic cusp (change in slope) at the training strain; data shown is an average of five replicates.  
}
\label{fig:protocol}
\end{figure}

In general, experiments are composed of two parts: training, in which several cycles of the same amplitude and type are applied, and readout, in which cycles of increasing amplitude are applied. Protocols for training and readout are depicted in Fig.~\ref{fig:protocol}(a,b). During readout, particle positions are measured so that an effective distance $\Delta_{min}$ can be calculated between any two configurations as described in the Methods section. This distance $\Delta_{min}$ is computed at the end of each cycle relative to the trained configuration measured at the beginning of the readout process (this trained configuration is shown schematically as a star in Fig.~\ref{fig:protocol}b). This provides a stroboscopic, or once-per-cycle, measure of the distance between the trained configuration and the configuration during readout, a quantity shown in both experimental and computational works to highlight a memory via a sudden increase in slope just after the training strain. %Such a curve is shown in Fig.~\ref{fig:protocol}c.

Two types of strain are used throughout. Strain type A, or $SA$, is almost exclusively $S1$ with only small contributions from $S2$ and compression. Strain type B, or $SB$, is primarily $S2$ and compression. The exact $S1$, $S2$, and compression values for these two strain types are shown in Fig.~\ref{fig:schematic}d. The magnitude of deformation for each type is taken to be the total shear strain $\sqrt{\gamma_1^2 + \gamma_2^2}$ and is taken to be negative when $S2$ is negative.

\begin{figure}
\centering
\includegraphics[width=8.6cm]{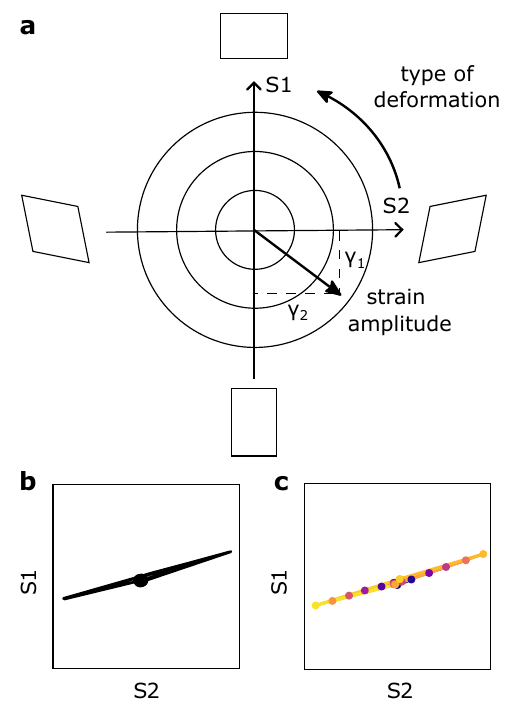}%
\caption{(a) Phase space of possible box shapes. The horizontal axis corresponds to strain of type $S2$ and the vertical axis corresponds to strain of type $S1$. The origin represents the original box shape, the total magnitude of strain is given by the distance from the origin $\sqrt{\gamma_1^2 + \gamma_2^2}$, and the polar angle describes the type of deformation applied. (b) Measured box shape in phase space during the same $SB$ training shown in Fig.~\ref{fig:protocol}b and (c) during the same $SB$ readout shown in Fig.~\ref{fig:protocol}c. In (c), color indicates early (dark purple) and late (light yellow) readout steps. Because the system traces back over the same box shapes several times, not all readout cycles are visible; dots have been added at the turning points to highlight the increasing cycle size shown explicitly in Fig.~\ref{fig:protocol}c.
}
\label{fig:polar}
\end{figure}

For the two-dimensional phase space of both $S1$ and $S2$ (or $SA$ and $SB$), the number of possible ways of training and reading out increases substantially. Rather than showing the training and readout as a function of time for both $SA$ and $SB$, it is convenient to represent them as a path through the two-dimensional phase space illustrated in Fig.~\ref{fig:polar}a. To provide some intuition for this space, Figs.~\ref{fig:polar}(b,c) show the same training and readout described in Fig.~\ref{fig:protocol}(a,b), keeping the convention that training cycles are shown in black and readout is shown in color with a transition from dark to light over the course of readout. 

\section*{Results}

When training and readout are performed along the same axis, the system exhibits a characteristic cusp in readout at the training amplitude as shown in Fig.~\ref{fig:protocol}c for $SB$. This cusp is a clear signature of a stored memory and is consistent both with other experimental realizations as well as with simulation results~\cite{fiocco2014encoding, keim2020global, benson2021memory}. A similar memory can be shown for a different training amplitude; for an untrained packing, readout produces no such cusp. Examples of both, as well similar plots for packings trained and read out with $SA$, are provided in the Supplementary Materials~\cite{SM}. 

%Fig.~\ref{fig:protocol}c shows the result of this process: a kink in the readout curve at the step corresponding to the cycle with amplitude equal to the training amplitude. 

\begin{comment}
Although this response is typical of amorphous solids, I note that our systems are extraordinarily small for such a well-resolved memory. It is therefore necessary to investigate other possible sources of the memory, in particular the box degrees of freedom as they are impacted by friction and hysteresis (which may provide a memory of their own) in the setup. 
\end{comment}

\begin{figure*}
\centering
\includegraphics[width=17.2cm]{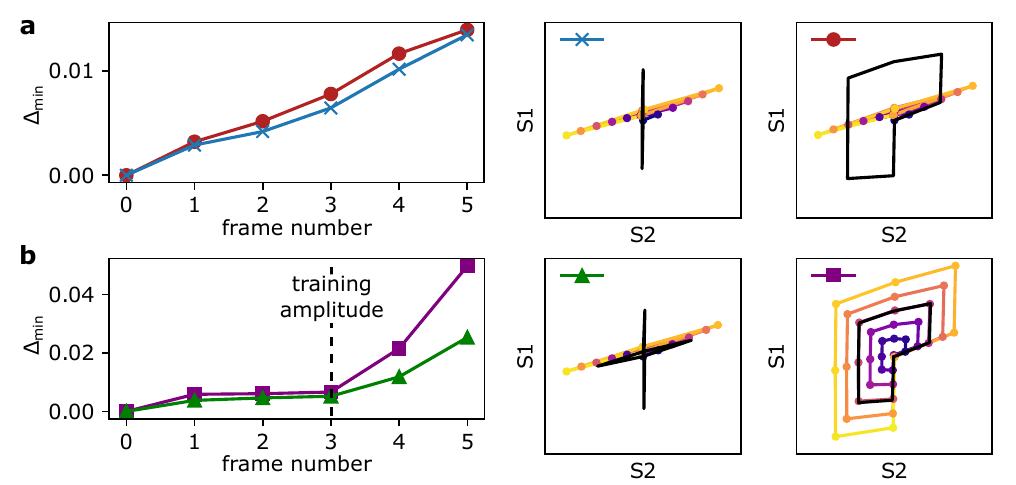}%
\caption{Left: Readout signal for different training and readout protocols. Right: phase space diagrams show the training (black) and readout (colorful) protocols used in each case. (a) No memory is observed for the cases where readout is done along $SB$ after training along $SA$ only (blue x's) or by tracing out a loop (red o's). 
(b) Memory is recovered when training is done alternately along $SB$ and $SA$ and readout along $SB$ (green triangles) or when training and readout are both done via loops in phase space (purple squares). $\Delta_{min}$ results are an average of between five and 10 replicates.
}
\label{fig:2d}
\end{figure*}

To explore combinations of both shear degrees of freedom, the scenario shown in Fig.~\ref{fig:protocol} is first modified only slightly, with training performed along $SA$ and readout performed along $SB$. Figure~\ref{fig:2d}a shows the readout in this case: a curve that increases monotonically with no clear cusp. Perhaps unsurprisingly, the memory of an amplitude also contains information about the direction in which training was applied, so that although a memory has formed it is not evident in an orthogonal readout. 

Second, the effect of alternating cycles along $SB$ and $SA$ is explored, beginning with $SB$ and ending with $SA$ to avoid the possibility that the system simply remembers the most recently applied deformation. Figure~\ref{fig:2d}b shows that the cusp in this case is restored: the system is able to learn the more complex, bi-directional training and hence the readout contains (partial) information about the training.

Finally, the order in which different parts of $SA$ and $SB$ cycles are applied during training is changed so that the path traced out in phase space encloses some area and the system does not return to its original (square box) configuration until the end of a full training cycle. Although one full cycle each of $SA$ and $SB$ are still applied over the course of a single training cycle, the readout signal, shown in Fig.~\ref{fig:2d}a, is notably different from that after training with ``unmixed'' alternating cycles of $SA$ and $SB$, with no signature of a memory. Reading out with cycles of the same type as those with which the system was trained restores the memory (Fig.~\ref{fig:2d}b).

\begin{figure*}
\centering
\includegraphics[width=17.2cm]{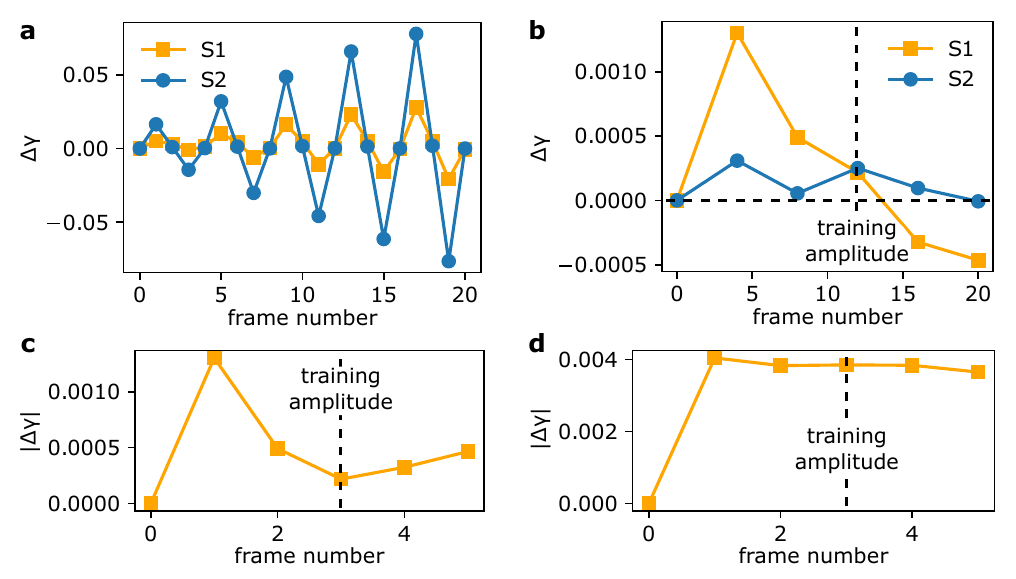}%
\caption{Measured strain values (a) throughout readout with applied shear type $SB$ (from the experiments reported in the main text Fig.~\ref{fig:protocol}c) and (b-d) at the beginning and end of every readout cycle. In all cases, the initial strain value is subtracted off so that the first data point is by definition 0: $\Delta \gamma = \gamma - \gamma_0$. (b) Same data as in (a), but only the stroboscopic values included (note the different y-axis scale: $S1$ and $S2$ vary only on the order of $0.1 \%$). $S1$ in particular returns closest to its initial strain value at the training amplitude. This is highlighted in (c), where only the absolute value of $S1$ is shown; this reveals a minimum at the training strain. The minimum is not present in (d), the corresponding plot for the experiments reported as green triangles in Fig.~\ref{fig:2d}b.}
\label{fig:box-shape}
\end{figure*}

\textit{Box effects} --- Because of friction and hysteresis in the mechanism that applies shear, the box itself does not come back to exactly the same shape after each cycle. In particular, the box itself appears to be trainable: after training, it returns to its original shape ($S1 = S2 = 0$) most closely at the point when the readout amplitude equals the training amplitude, as shown in Fig.~\ref{fig:box-shape}. This is highlighted by showing only stroboscopic measurements, or measurements taken just once per cycle, as in Figs.~\ref{fig:box-shape}(b-d). These are all cases where the box shape is nominally its original shape, yet these measurements reveal a small residual strain ($\sim 50$ times smaller than the applied training cycles). Residual stroboscopic $S2$ shows no particular form, yet $S1$ returns closest to its original value when the readout amplitude reaches the training amplitude. This is seen more clearly when the absolute value of $S1$ is plotted, revealing a minimum at the training amplitude. 

Because this box memory takes on a form similar to that of the particle memory shown in Figs.~\ref{fig:protocol} and~\ref{fig:2d}, one can ask whether the apparent particle memory is actually due to the box memory. It is unclear how to estimate the relative strengths of the two effects, making a quantitative analysis difficult, yet there are cases like the one shown in Fig.~\ref{fig:box-shape}d where box memory disappears yet the corresponding particle memory (shown in Fig.~\ref{fig:2d}b) exists; other such examples are shown in the Supplemental Materials. While the box memory may account for some of the memory seen during readout, evidently much of the memory is also encoded in the particle configurations. 

\section*{Discussion}

%\textit{Discussion} --- 

In most simulations, particles are taken to be perfect disks or spheres so that the only degrees of freedom in which memory can be stored are the particle positions. In these experiments, there exist a number of additional degrees of freedom per particle, both macroscopic (as in the out-of-plane rotations of individual hydrogel spheres) and microscopic (frictional contacts and sliding). As noted above, the box degrees of freedom may also play a role. 
The presence of memory in both the particles and the box highlights the pervasiveness of memory in a much wider range of physical systems. 
%In addition, the box degrees of freedom themselves possess a memory so that, during readout, the box will often return back to its initial shape most exactly at the training strain. As the box shape is used in analysis (see Methods), this likely contributes to the cusps observed in readout signal. However, note that cusps can be observed even in the absence of such a box memory, suggesting that particle positions are indeed responsible for at least some component of the observed effect; see the Supplemental Materials for more information~\cite{SM}.

Though the effect of driving timescale on memory is outside the scope of this work, the system studied here includes several features of interest in terms of dynamics. Hydrogel spheres are relatively low-friction particles compared with, for example, rough spheres or disks. However, detailed work has shown that hydrogels exhibit complex time-dependent frictional interactions even with smooth surfaces, with relaxation occurring on minutes or even hours-long timescales~\cite{cuccia2020pore}. Combined with an easily modified actuation speed, this makes it possible to explore the interplay between inherent relaxation timescales and driving frequency. In particular, it would be interesting to see an experimental study of how this affects trainability, a question so far primarily considered in model systems~\cite{lindeman2023competition}. 

Given that the particles are open to air, the particle radii will shrink over hours- or days-long timescales as water evaporates. With a clever experimental design, it may be possible to exhibit some control over these radii, for example by restoring enough fluid to conserve the total volume over time. If these radii change preferentially depending on (for example) local pressure or exposed surface area, they may become additional degrees of freedom, a feature which can allow for the formation of extremely stable configurations~\cite{hagh2022transient}.

The focus of this work has been on the response of the system during readout with a fixed training time (five cycles). However, there is also the question of training: how long does it take to train in steady-state behavior as a function of the complexity of the cycle? Can more complex memories be trained in, for example different shear amplitudes in $S1$ and $S2$ or even adding compression as an independent axis? Do new kinds of limit cycles emerge, as was shown in a simple model system under periodic, two-dimensional driving~\cite{hexner2015self}?

Even a single shear degree of freedom offers a large space of possible driving in the sense that training can be done at any amplitude, or for multiple amplitudes in different orders. In this case, the question is often one of the memory capacity of a system~\cite{keim2022mechanical}. A single readout protocol is expected to reveal all of the stored memories. Here, the data show that when training can be performed along more than one axis, the choice of readout protocol becomes a crucial one. In particular, it becomes important to distinguish between \textit{forming} a memory (as evidenced, for example, by the arrival at a steady state during driving) and \textit{observing} that memory through a particular choice of readout protocol. Most readout protocols are destructive: the act of reading out a memory erases it because the large-amplitude cycles wipe out most information about the previous training~\cite{lindeman2021multiple, lindeman2023isolating}. Further experimental and theoretical studies are necessary to determine whether there exist more general protocols capable of revealing the memory of a full path through phase space. 

\section*{Methods}

%\textit{Methods} --- 

Hydrogel spheres (Crystal Accents Deco Beads from JRM Chemical, Inc.), swelled for at least 12 hours with tap water, are used as particles. Packings of hydrogels are sheared by a custom-built setup, shown in detail in the Supplementary Materials~\cite{SM}, that allows $SA$ and $SB$ to be independently controlled using two feedback rod linear actuators (Firgelli Automations FA-PO-35-12-2), two motor drivers (Hiletego BTS7960 43A), and an Arduino Uno. 

Before each experiment, between 34 and 37 particles are loaded into the shear cell by hand, with an effort made to ensure some polydispersity to suppress crystallization. The number of particles is in a range intended to keep the system jammed throughout the shear cycle but avoid three-dimensional buckling (escape to the third dimension). Experiments with buckling events are not excluded as long as bucklers come back into the plane at the end of each cycle.

Particles are imaged throughout the shear cycle using a Nikon D90 and the resulting videos are analyzed in ImageJ. This allows for simultaneous tracking of the boundary conditions and the particle positions as detailed below.

\textit{Boundary conditions:} Any bounding parallelogram can be defined by two vectors $\vec{a}$ and $\vec{b}$, which can be put into a matrix $B = [ \vec{a} \hspace{1mm} \vec{b} ]$. In this notation, for example, $B$ corresponding to shear type $S2$ is given by 
$$B_2 = \alpha \begin{bmatrix}
1  && \gamma \\ \gamma && 1 
\end{bmatrix},$$ 
where $\alpha = (1-\gamma^2)^{-1/2}$ enforces constant area. The direction of extension (in this case, along the horizontal) can be rotated by an angle $\theta$ by applying the rotation matrix $R(\theta)$:
$$R(\theta) B_2 R^{-1}(\theta) = \begin{bmatrix}
1 + \gamma \sin{2 \theta} && \gamma \cos{2 \theta} \\ \gamma \cos{2 \theta} && 1 - \gamma \sin{2 \theta}
\end{bmatrix}.$$
In the case that $\theta = \pi/4$, this becomes exactly 
$$R(\pi/4) B_1 R^{-1}(\pi/4) = \alpha \begin{bmatrix}
1 + \gamma && 0 \\ 0 && 1 - \gamma
\end{bmatrix} = B_1,$$
the matrix $B$ corresponding to shear type $S1$. There is thus an equivalence between this ``polar'' coordinate system, in which $2\theta$ ranges from 0 to $2 \pi$ and $\gamma$ describes the total strain in the system, and a ``Cartesian'' coordinate system in which I define strains corresponding to $S1$ and $S2$ as $\gamma_1 = \gamma \sin{2\theta}$ and $\gamma_2=\gamma \cos{2\theta}$, respectively. Any box deformation is then describable as a linear combination of $S1$ and $S2$, with total strain given by $\gamma = \sqrt{\gamma_1^2 + \gamma_2^2}$ and type of strain given by the signs and relative magnitudes of $\gamma_1$ and $\gamma_2$. This equivalence is shown schematically in Fig.~\ref{fig:polar}a.

\textit{Tracking:} Boundaries of the shear cell are marked with red and blue dots (on the top/bottom and left/right walls, respectively). Dots are fit to ellipses using the built-in ImageJ ``Analyze Particles'' function and the ellipse centroid positions are fit to a parallelogram in python. This parallelogram provides full information about the amplitudes of $S1$ and $S2$ as well as any compression relative to the initial frame of a given experiment.

To extract the particle positions, the image is thresholded and a watershed algorithm applied, followed by two iterations of ``erosion'' (removing the outer layer of each particle) to separate out individual particles. Particles are again fit to ellipses with ImageJ's built-in ``Analyze Particles'' function and the positions of each ellipse's center are exported for analysis in python.

\textit{Analysis:} Particles are linked from frame to frame using a simple distance limit, linking any given particle in frame $i$ to the nearest particle in frame $i+1$ to get a per-particle distance $\delta_{min}$. The ``distance'' between configurations $\Delta_{min}$ is then given by the average of this minimum change in position:
$$ \Delta_{min} = \frac{1}{N} \sum \delta_{min}$$
In the case that particles do not travel farther than the typical radius from one cycle to the next, this corresponds to exactly the average particle displacement; if particles move much farther than one radius this measure becomes a lower bound on the actual distance travelled (typically still quite large as particles are unlikely to exactly swap positions). Particles are linked from frame-to-frame with nearly $100 \%$ fidelity, allowing for the calculation of system-wide displacement and hence the $\Delta_{min}$ values reported in Figs.~\ref{fig:protocol} and~\ref{fig:2d}. 

To mitigate small changes in boundary conditions from cycle to cycle, positions are mapped from pixel space into a 1x1 box using the inverse of the boundary conditions $B$ as described below. The mean particle motion is subtracted before $\Delta_{min}$ is calculated.

Given the best-fit box shape $B_{fit}$, arbitrary positions $(x, y)$ in a square box transform via affine deformation to their corresponding positions $(x', y')$ in a sheared box as 
$$\begin{bmatrix}
x'\\ y'
\end{bmatrix} = 
B_{fit}
\begin{bmatrix}
x\\ y
\end{bmatrix}.$$ It follows that affine motion can be removed from measured particle positions $(x_m, y_m)$ by instead using the position mapped back into a one-by-one square box
$$\begin{bmatrix}
x_{square}\\ y_{square}
\end{bmatrix} = 
B_{fit}^{-1}
\begin{bmatrix}
x_m\\ y_m
\end{bmatrix}.$$

For a generic parallelogram boundary condition, $B$ can be written in the ``cartesian'' strain coordinates $\gamma_1$ and $\gamma_2$ as
$$
\text{B} = 
\beta
\begin{bmatrix}
1 + \gamma_1 & \gamma_2\\
\gamma_2 & 1 - \gamma_1
\end{bmatrix},
%\label{eqn:LV}
$$
where $\gamma_1$ is the amplitude of $S1$, $\gamma_2$ is the amplitude of $S2$, and $\beta$ is related to the compression.
\begin{comment}
\begin{equation}
\text{B} = 
\frac{1}{\sqrt{\gamma_c}}
\frac{1}{\sqrt{1 - \gamma_1^2 - \gamma_2^2}}
\begin{bmatrix}
1 + \gamma_1 & \gamma_2\\
\gamma_2 & 1 - \gamma_1
\end{bmatrix},
\label{eqn:LV}
\end{equation}
where $\gamma_1$ is the amplitude of $S1$, $\gamma_2$ is the amplitude of $S2$, and I have explicitly taken into account the changes in volume due to $\gamma_1$ and $\gamma_2$ such that $\gamma_c$ represents the fractional compression relative to the unstrained case. 
\end{comment}
Because rotations of the entire system are irrelevant, the fitted parallelogram $B_{fit}$ is rotated so that it is symmetric (to match the form of $B$ given above) before fitting for $\gamma_1$, $\gamma_2$, and $\gamma_c$ relative to the first frame of the experiment. 

\section*{Acknowledgements}

I thank Samar Alqatari, Justin Burton, Savannah Gowen, Nathan Keim, Zhaoning Liu, Ayanna Matthews, Nidhi Pashine, Baudouin Saintyves, and Khá-Î Tô. Sidney Nagel in particular was central not only to the development of ideas presented here but to my own development as a scientist. This work was supported by the NSF MRSEC program NSF-DMR 2011854. C.W.L. was supported in part by NSF Graduate Research Fellowship Grant DGE-1746045.

\bibliography{apssamp}

\begin{thebibliography}{41}%
\makeatletter
\providecommand \@ifxundefined [1]{%
 \@ifx{#1\undefined}
}%
\providecommand \@ifnum [1]{%
 \ifnum #1\expandafter \@firstoftwo
 \else \expandafter \@secondoftwo
 \fi
}%
\providecommand \@ifx [1]{%
 \ifx #1\expandafter \@firstoftwo
 \else \expandafter \@secondoftwo
 \fi
}%
\providecommand \natexlab [1]{#1}%
\providecommand \enquote  [1]{``#1''}%
\providecommand \bibnamefont  [1]{#1}%
\providecommand \bibfnamefont [1]{#1}%
\providecommand \citenamefont [1]{#1}%
\providecommand \href@noop [0]{\@secondoftwo}%
\providecommand \href [0]{\begingroup \@sanitize@url \@href}%
\providecommand \@href[1]{\@@startlink{#1}\@@href}%
\providecommand \@@href[1]{\endgroup#1\@@endlink}%
\providecommand \@sanitize@url [0]{\catcode `\\12\catcode `\$12\catcode `\&12\catcode `\#12\catcode `\^12\catcode `\_12\catcode `\%12\relax}%
\providecommand \@@startlink[1]{}%
\providecommand \@@endlink[0]{}%
\providecommand \url  [0]{\begingroup\@sanitize@url \@url }%
\providecommand \@url [1]{\endgroup\@href {#1}{\urlprefix }}%
\providecommand \urlprefix  [0]{URL }%
\providecommand \Eprint [0]{\href }%
\providecommand \doibase [0]{http://dx.doi.org/}%
\providecommand \selectlanguage [0]{\@gobble}%
\providecommand \bibinfo  [0]{\@secondoftwo}%
\providecommand \bibfield  [0]{\@secondoftwo}%
\providecommand \translation [1]{[#1]}%
\providecommand \BibitemOpen [0]{}%
\providecommand \bibitemStop [0]{}%
\providecommand \bibitemNoStop [0]{.\EOS\space}%
\providecommand \EOS [0]{\spacefactor3000\relax}%
\providecommand \BibitemShut  [1]{\csname bibitem#1\endcsname}%
\let\auto@bib@innerbib\@empty
%</preamble>
\bibitem [{\citenamefont {Toiya}\ \emph {et~al.}(2004)\citenamefont {Toiya}, \citenamefont {Stambaugh},\ and\ \citenamefont {Losert}}]{toiya2004transient}%
  \BibitemOpen
  \bibfield  {author} {\bibinfo {author} {\bibfnamefont {M.}~\bibnamefont {Toiya}}, \bibinfo {author} {\bibfnamefont {J.}~\bibnamefont {Stambaugh}}, \ and\ \bibinfo {author} {\bibfnamefont {W.}~\bibnamefont {Losert}},\ }\href@noop {} {\bibfield  {journal} {\bibinfo  {journal} {Physical review letters}\ }\textbf {\bibinfo {volume} {93}},\ \bibinfo {pages} {088001} (\bibinfo {year} {2004})}\BibitemShut {NoStop}%
\bibitem [{\citenamefont {Gadala-Maria}\ and\ \citenamefont {Acrivos}(1980)}]{gadala1980shear}%
  \BibitemOpen
  \bibfield  {author} {\bibinfo {author} {\bibfnamefont {F.}~\bibnamefont {Gadala-Maria}}\ and\ \bibinfo {author} {\bibfnamefont {A.}~\bibnamefont {Acrivos}},\ }\href@noop {} {\bibfield  {journal} {\bibinfo  {journal} {Journal of rheology}\ }\textbf {\bibinfo {volume} {24}},\ \bibinfo {pages} {799} (\bibinfo {year} {1980})}\BibitemShut {NoStop}%
\bibitem [{\citenamefont {Mukherji}\ \emph {et~al.}(2019)\citenamefont {Mukherji}, \citenamefont {Kandula}, \citenamefont {Sood},\ and\ \citenamefont {Ganapathy}}]{mukherji2019strength}%
  \BibitemOpen
  \bibfield  {author} {\bibinfo {author} {\bibfnamefont {S.}~\bibnamefont {Mukherji}}, \bibinfo {author} {\bibfnamefont {N.}~\bibnamefont {Kandula}}, \bibinfo {author} {\bibfnamefont {A.}~\bibnamefont {Sood}}, \ and\ \bibinfo {author} {\bibfnamefont {R.}~\bibnamefont {Ganapathy}},\ }\href@noop {} {\bibfield  {journal} {\bibinfo  {journal} {Physical review letters}\ }\textbf {\bibinfo {volume} {122}},\ \bibinfo {pages} {158001} (\bibinfo {year} {2019})}\BibitemShut {NoStop}%
\bibitem [{\citenamefont {Keim}\ \emph {et~al.}(2020)\citenamefont {Keim}, \citenamefont {Hass}, \citenamefont {Kroger},\ and\ \citenamefont {Wieker}}]{keim2020global}%
  \BibitemOpen
  \bibfield  {author} {\bibinfo {author} {\bibfnamefont {N.~C.}\ \bibnamefont {Keim}}, \bibinfo {author} {\bibfnamefont {J.}~\bibnamefont {Hass}}, \bibinfo {author} {\bibfnamefont {B.}~\bibnamefont {Kroger}}, \ and\ \bibinfo {author} {\bibfnamefont {D.}~\bibnamefont {Wieker}},\ }\href@noop {} {\bibfield  {journal} {\bibinfo  {journal} {Physical Review Research}\ }\textbf {\bibinfo {volume} {2}},\ \bibinfo {pages} {012004} (\bibinfo {year} {2020})}\BibitemShut {NoStop}%
\bibitem [{\citenamefont {Candela}(2023)}]{candela2023complex}%
  \BibitemOpen
  \bibfield  {author} {\bibinfo {author} {\bibfnamefont {D.}~\bibnamefont {Candela}},\ }\href@noop {} {\bibfield  {journal} {\bibinfo  {journal} {Physical Review Letters}\ }\textbf {\bibinfo {volume} {130}},\ \bibinfo {pages} {268202} (\bibinfo {year} {2023})}\BibitemShut {NoStop}%
\bibitem [{\citenamefont {Keim}\ \emph {et~al.}(2019)\citenamefont {Keim}, \citenamefont {Paulsen}, \citenamefont {Zeravcic}, \citenamefont {Sastry},\ and\ \citenamefont {Nagel}}]{keim2019memory}%
  \BibitemOpen
  \bibfield  {author} {\bibinfo {author} {\bibfnamefont {N.~C.}\ \bibnamefont {Keim}}, \bibinfo {author} {\bibfnamefont {J.~D.}\ \bibnamefont {Paulsen}}, \bibinfo {author} {\bibfnamefont {Z.}~\bibnamefont {Zeravcic}}, \bibinfo {author} {\bibfnamefont {S.}~\bibnamefont {Sastry}}, \ and\ \bibinfo {author} {\bibfnamefont {S.~R.}\ \bibnamefont {Nagel}},\ }\href@noop {} {\bibfield  {journal} {\bibinfo  {journal} {Reviews of Modern Physics}\ }\textbf {\bibinfo {volume} {91}},\ \bibinfo {pages} {035002} (\bibinfo {year} {2019})}\BibitemShut {NoStop}%
\bibitem [{\citenamefont {Paulsen}\ and\ \citenamefont {Keim}(2024)}]{paulsen2024mechanical}%
  \BibitemOpen
  \bibfield  {author} {\bibinfo {author} {\bibfnamefont {J.~D.}\ \bibnamefont {Paulsen}}\ and\ \bibinfo {author} {\bibfnamefont {N.~C.}\ \bibnamefont {Keim}},\ }\href@noop {} {\bibfield  {journal} {\bibinfo  {journal} {arXiv preprint arXiv:2405.08158}\ } (\bibinfo {year} {2024})}\BibitemShut {NoStop}%
\bibitem [{\citenamefont {Fiocco}\ \emph {et~al.}(2014)\citenamefont {Fiocco}, \citenamefont {Foffi},\ and\ \citenamefont {Sastry}}]{fiocco2014encoding}%
  \BibitemOpen
  \bibfield  {author} {\bibinfo {author} {\bibfnamefont {D.}~\bibnamefont {Fiocco}}, \bibinfo {author} {\bibfnamefont {G.}~\bibnamefont {Foffi}}, \ and\ \bibinfo {author} {\bibfnamefont {S.}~\bibnamefont {Sastry}},\ }\href@noop {} {\bibfield  {journal} {\bibinfo  {journal} {Physical review letters}\ }\textbf {\bibinfo {volume} {112}},\ \bibinfo {pages} {025702} (\bibinfo {year} {2014})}\BibitemShut {NoStop}%
\bibitem [{\citenamefont {Mungan}\ and\ \citenamefont {Witten}(2019)}]{mungan2019cyclic}%
  \BibitemOpen
  \bibfield  {author} {\bibinfo {author} {\bibfnamefont {M.}~\bibnamefont {Mungan}}\ and\ \bibinfo {author} {\bibfnamefont {T.~A.}\ \bibnamefont {Witten}},\ }\href@noop {} {\bibfield  {journal} {\bibinfo  {journal} {Physical Review E}\ }\textbf {\bibinfo {volume} {99}},\ \bibinfo {pages} {052132} (\bibinfo {year} {2019})}\BibitemShut {NoStop}%
\bibitem [{\citenamefont {Keim}\ and\ \citenamefont {Paulsen}(2021)}]{keim2021multiperiodic}%
  \BibitemOpen
  \bibfield  {author} {\bibinfo {author} {\bibfnamefont {N.~C.}\ \bibnamefont {Keim}}\ and\ \bibinfo {author} {\bibfnamefont {J.~D.}\ \bibnamefont {Paulsen}},\ }\href@noop {} {\bibfield  {journal} {\bibinfo  {journal} {Science Advances}\ }\textbf {\bibinfo {volume} {7}},\ \bibinfo {pages} {eabg7685} (\bibinfo {year} {2021})}\BibitemShut {NoStop}%
\bibitem [{\citenamefont {Szulc}\ \emph {et~al.}(2022)\citenamefont {Szulc}, \citenamefont {Mungan},\ and\ \citenamefont {Regev}}]{szulc2022cooperative}%
  \BibitemOpen
  \bibfield  {author} {\bibinfo {author} {\bibfnamefont {A.}~\bibnamefont {Szulc}}, \bibinfo {author} {\bibfnamefont {M.}~\bibnamefont {Mungan}}, \ and\ \bibinfo {author} {\bibfnamefont {I.}~\bibnamefont {Regev}},\ }\href@noop {} {\bibfield  {journal} {\bibinfo  {journal} {The Journal of Chemical Physics}\ }\textbf {\bibinfo {volume} {156}} (\bibinfo {year} {2022})}\BibitemShut {NoStop}%
\bibitem [{\citenamefont {Kumar}\ \emph {et~al.}(2022)\citenamefont {Kumar}, \citenamefont {Patinet}, \citenamefont {Maloney}, \citenamefont {Regev}, \citenamefont {Vandembroucq},\ and\ \citenamefont {Mungan}}]{kumar2022mapping}%
  \BibitemOpen
  \bibfield  {author} {\bibinfo {author} {\bibfnamefont {D.}~\bibnamefont {Kumar}}, \bibinfo {author} {\bibfnamefont {S.}~\bibnamefont {Patinet}}, \bibinfo {author} {\bibfnamefont {C.~E.}\ \bibnamefont {Maloney}}, \bibinfo {author} {\bibfnamefont {I.}~\bibnamefont {Regev}}, \bibinfo {author} {\bibfnamefont {D.}~\bibnamefont {Vandembroucq}}, \ and\ \bibinfo {author} {\bibfnamefont {M.}~\bibnamefont {Mungan}},\ }\href@noop {} {\bibfield  {journal} {\bibinfo  {journal} {The Journal of Chemical Physics}\ }\textbf {\bibinfo {volume} {157}} (\bibinfo {year} {2022})}\BibitemShut {NoStop}%
\bibitem [{\citenamefont {Movsheva}\ and\ \citenamefont {Witten}(2023)}]{movsheva2023granular}%
  \BibitemOpen
  \bibfield  {author} {\bibinfo {author} {\bibfnamefont {A.}~\bibnamefont {Movsheva}}\ and\ \bibinfo {author} {\bibfnamefont {T.~A.}\ \bibnamefont {Witten}},\ }\href@noop {} {\bibfield  {journal} {\bibinfo  {journal} {The European Physical Journal E}\ }\textbf {\bibinfo {volume} {46}},\ \bibinfo {pages} {84} (\bibinfo {year} {2023})}\BibitemShut {NoStop}%
\bibitem [{\citenamefont {Royer}\ and\ \citenamefont {Chaikin}(2015)}]{royer2015precisely}%
  \BibitemOpen
  \bibfield  {author} {\bibinfo {author} {\bibfnamefont {J.~R.}\ \bibnamefont {Royer}}\ and\ \bibinfo {author} {\bibfnamefont {P.~M.}\ \bibnamefont {Chaikin}},\ }\href@noop {} {\bibfield  {journal} {\bibinfo  {journal} {Proceedings of the National Academy of Sciences}\ }\textbf {\bibinfo {volume} {112}},\ \bibinfo {pages} {49} (\bibinfo {year} {2015})}\BibitemShut {NoStop}%
\bibitem [{\citenamefont {Mungan}\ \emph {et~al.}(2019)\citenamefont {Mungan}, \citenamefont {Sastry}, \citenamefont {Dahmen},\ and\ \citenamefont {Regev}}]{mungan2019networks}%
  \BibitemOpen
  \bibfield  {author} {\bibinfo {author} {\bibfnamefont {M.}~\bibnamefont {Mungan}}, \bibinfo {author} {\bibfnamefont {S.}~\bibnamefont {Sastry}}, \bibinfo {author} {\bibfnamefont {K.}~\bibnamefont {Dahmen}}, \ and\ \bibinfo {author} {\bibfnamefont {I.}~\bibnamefont {Regev}},\ }\href@noop {} {\bibfield  {journal} {\bibinfo  {journal} {Physical review letters}\ }\textbf {\bibinfo {volume} {123}},\ \bibinfo {pages} {178002} (\bibinfo {year} {2019})}\BibitemShut {NoStop}%
\bibitem [{\citenamefont {Benson}\ \emph {et~al.}(2021)\citenamefont {Benson}, \citenamefont {Peshkov}, \citenamefont {Richardson},\ and\ \citenamefont {Losert}}]{benson2021memory}%
  \BibitemOpen
  \bibfield  {author} {\bibinfo {author} {\bibfnamefont {Z.~A.}\ \bibnamefont {Benson}}, \bibinfo {author} {\bibfnamefont {A.}~\bibnamefont {Peshkov}}, \bibinfo {author} {\bibfnamefont {D.~C.}\ \bibnamefont {Richardson}}, \ and\ \bibinfo {author} {\bibfnamefont {W.}~\bibnamefont {Losert}},\ }\href@noop {} {\bibfield  {journal} {\bibinfo  {journal} {Physical Review E}\ }\textbf {\bibinfo {volume} {103}},\ \bibinfo {pages} {062906} (\bibinfo {year} {2021})}\BibitemShut {NoStop}%
\bibitem [{\citenamefont {Schwen}\ \emph {et~al.}(2020)\citenamefont {Schwen}, \citenamefont {Ramaswamy}, \citenamefont {Cheng}, \citenamefont {Jan},\ and\ \citenamefont {Cohen}}]{schwen2020embedding}%
  \BibitemOpen
  \bibfield  {author} {\bibinfo {author} {\bibfnamefont {E.~M.}\ \bibnamefont {Schwen}}, \bibinfo {author} {\bibfnamefont {M.}~\bibnamefont {Ramaswamy}}, \bibinfo {author} {\bibfnamefont {C.-M.}\ \bibnamefont {Cheng}}, \bibinfo {author} {\bibfnamefont {L.}~\bibnamefont {Jan}}, \ and\ \bibinfo {author} {\bibfnamefont {I.}~\bibnamefont {Cohen}},\ }\href@noop {} {\bibfield  {journal} {\bibinfo  {journal} {Soft matter}\ }\textbf {\bibinfo {volume} {16}},\ \bibinfo {pages} {3746} (\bibinfo {year} {2020})}\BibitemShut {NoStop}%
\bibitem [{\citenamefont {Galloway}\ \emph {et~al.}(2022)\citenamefont {Galloway}, \citenamefont {Teich}, \citenamefont {Ma}, \citenamefont {Kammer}, \citenamefont {Graham}, \citenamefont {Keim}, \citenamefont {Reina}, \citenamefont {Jerolmack}, \citenamefont {Yodh},\ and\ \citenamefont {Arratia}}]{galloway2022relationships}%
  \BibitemOpen
  \bibfield  {author} {\bibinfo {author} {\bibfnamefont {K.}~\bibnamefont {Galloway}}, \bibinfo {author} {\bibfnamefont {E.}~\bibnamefont {Teich}}, \bibinfo {author} {\bibfnamefont {X.}~\bibnamefont {Ma}}, \bibinfo {author} {\bibfnamefont {C.}~\bibnamefont {Kammer}}, \bibinfo {author} {\bibfnamefont {I.}~\bibnamefont {Graham}}, \bibinfo {author} {\bibfnamefont {N.}~\bibnamefont {Keim}}, \bibinfo {author} {\bibfnamefont {C.}~\bibnamefont {Reina}}, \bibinfo {author} {\bibfnamefont {D.}~\bibnamefont {Jerolmack}}, \bibinfo {author} {\bibfnamefont {A.}~\bibnamefont {Yodh}}, \ and\ \bibinfo {author} {\bibfnamefont {P.}~\bibnamefont {Arratia}},\ }\href@noop {} {\bibfield  {journal} {\bibinfo  {journal} {Nature Physics}\ }\textbf {\bibinfo {volume} {18}},\ \bibinfo {pages} {565} (\bibinfo {year} {2022})}\BibitemShut {NoStop}%
\bibitem [{\citenamefont {Chen}\ \emph {et~al.}(2023)\citenamefont {Chen}, \citenamefont {Rogers}, \citenamefont {Narayanan}, \citenamefont {Harden},\ and\ \citenamefont {Leheny}}]{chen2023microstructural}%
  \BibitemOpen
  \bibfield  {author} {\bibinfo {author} {\bibfnamefont {Y.}~\bibnamefont {Chen}}, \bibinfo {author} {\bibfnamefont {S.~A.}\ \bibnamefont {Rogers}}, \bibinfo {author} {\bibfnamefont {S.}~\bibnamefont {Narayanan}}, \bibinfo {author} {\bibfnamefont {J.~L.}\ \bibnamefont {Harden}}, \ and\ \bibinfo {author} {\bibfnamefont {R.~L.}\ \bibnamefont {Leheny}},\ }\href@noop {} {\bibfield  {journal} {\bibinfo  {journal} {arXiv preprint arXiv:2312.17696}\ } (\bibinfo {year} {2023})}\BibitemShut {NoStop}%
\bibitem [{\citenamefont {Morse}\ \emph {et~al.}(2020)\citenamefont {Morse}, \citenamefont {Wijtmans}, \citenamefont {Van~Deen}, \citenamefont {Van~Hecke},\ and\ \citenamefont {Manning}}]{morse2020differences}%
  \BibitemOpen
  \bibfield  {author} {\bibinfo {author} {\bibfnamefont {P.}~\bibnamefont {Morse}}, \bibinfo {author} {\bibfnamefont {S.}~\bibnamefont {Wijtmans}}, \bibinfo {author} {\bibfnamefont {M.}~\bibnamefont {Van~Deen}}, \bibinfo {author} {\bibfnamefont {M.}~\bibnamefont {Van~Hecke}}, \ and\ \bibinfo {author} {\bibfnamefont {M.~L.}\ \bibnamefont {Manning}},\ }\href@noop {} {\bibfield  {journal} {\bibinfo  {journal} {Physical Review Research}\ }\textbf {\bibinfo {volume} {2}},\ \bibinfo {pages} {023179} (\bibinfo {year} {2020})}\BibitemShut {NoStop}%
\bibitem [{\citenamefont {Lindeman}\ and\ \citenamefont {Nagel}(2024)}]{lindeman2024minimal}%
  \BibitemOpen
  \bibfield  {author} {\bibinfo {author} {\bibfnamefont {C.~W.}\ \bibnamefont {Lindeman}}\ and\ \bibinfo {author} {\bibfnamefont {S.~R.}\ \bibnamefont {Nagel}},\ }\href@noop {} {\bibfield  {journal} {\bibinfo  {journal} {arXiv preprint arXiv:2403.01679}\ } (\bibinfo {year} {2024})}\BibitemShut {NoStop}%
\bibitem [{\citenamefont {Xu}\ \emph {et~al.}(2017)\citenamefont {Xu}, \citenamefont {Liu},\ and\ \citenamefont {Nagel}}]{xu2017instabilities}%
  \BibitemOpen
  \bibfield  {author} {\bibinfo {author} {\bibfnamefont {N.}~\bibnamefont {Xu}}, \bibinfo {author} {\bibfnamefont {A.~J.}\ \bibnamefont {Liu}}, \ and\ \bibinfo {author} {\bibfnamefont {S.~R.}\ \bibnamefont {Nagel}},\ }\href@noop {} {\bibfield  {journal} {\bibinfo  {journal} {Physical review letters}\ }\textbf {\bibinfo {volume} {119}},\ \bibinfo {pages} {215502} (\bibinfo {year} {2017})}\BibitemShut {NoStop}%
\bibitem [{\citenamefont {Xu}\ \emph {et~al.}(2023)\citenamefont {Xu}, \citenamefont {Zhang}, \citenamefont {Liu}, \citenamefont {Nagel},\ and\ \citenamefont {Xu}}]{xu2023discontinuous}%
  \BibitemOpen
  \bibfield  {author} {\bibinfo {author} {\bibfnamefont {D.}~\bibnamefont {Xu}}, \bibinfo {author} {\bibfnamefont {S.}~\bibnamefont {Zhang}}, \bibinfo {author} {\bibfnamefont {A.~J.}\ \bibnamefont {Liu}}, \bibinfo {author} {\bibfnamefont {S.~R.}\ \bibnamefont {Nagel}}, \ and\ \bibinfo {author} {\bibfnamefont {N.}~\bibnamefont {Xu}},\ }\href@noop {} {\bibfield  {journal} {\bibinfo  {journal} {Proceedings of the National Academy of Sciences}\ }\textbf {\bibinfo {volume} {120}},\ \bibinfo {pages} {e2304974120} (\bibinfo {year} {2023})}\BibitemShut {NoStop}%
\bibitem [{\citenamefont {Richard}\ \emph {et~al.}(2020)\citenamefont {Richard}, \citenamefont {Ozawa}, \citenamefont {Patinet}, \citenamefont {Stanifer}, \citenamefont {Shang}, \citenamefont {Ridout}, \citenamefont {Xu}, \citenamefont {Zhang}, \citenamefont {Morse}, \citenamefont {Barrat} \emph {et~al.}}]{richard2020predicting}%
  \BibitemOpen
  \bibfield  {author} {\bibinfo {author} {\bibfnamefont {D.}~\bibnamefont {Richard}}, \bibinfo {author} {\bibfnamefont {M.}~\bibnamefont {Ozawa}}, \bibinfo {author} {\bibfnamefont {S.}~\bibnamefont {Patinet}}, \bibinfo {author} {\bibfnamefont {E.}~\bibnamefont {Stanifer}}, \bibinfo {author} {\bibfnamefont {B.}~\bibnamefont {Shang}}, \bibinfo {author} {\bibfnamefont {S.}~\bibnamefont {Ridout}}, \bibinfo {author} {\bibfnamefont {B.}~\bibnamefont {Xu}}, \bibinfo {author} {\bibfnamefont {G.}~\bibnamefont {Zhang}}, \bibinfo {author} {\bibfnamefont {P.}~\bibnamefont {Morse}}, \bibinfo {author} {\bibfnamefont {J.-L.}\ \bibnamefont {Barrat}},  \emph {et~al.},\ }\href@noop {} {\bibfield  {journal} {\bibinfo  {journal} {Physical Review Materials}\ }\textbf {\bibinfo {volume} {4}},\ \bibinfo {pages} {113609} (\bibinfo {year} {2020})}\BibitemShut {NoStop}%
\bibitem [{\citenamefont {Manning}\ and\ \citenamefont {Liu}(2011)}]{manning2011vibrational}%
  \BibitemOpen
  \bibfield  {author} {\bibinfo {author} {\bibfnamefont {M.~L.}\ \bibnamefont {Manning}}\ and\ \bibinfo {author} {\bibfnamefont {A.~J.}\ \bibnamefont {Liu}},\ }\href@noop {} {\bibfield  {journal} {\bibinfo  {journal} {Physical Review Letters}\ }\textbf {\bibinfo {volume} {107}},\ \bibinfo {pages} {108302} (\bibinfo {year} {2011})}\BibitemShut {NoStop}%
\bibitem [{\citenamefont {Cubuk}\ \emph {et~al.}(2015)\citenamefont {Cubuk}, \citenamefont {Schoenholz}, \citenamefont {Rieser}, \citenamefont {Malone}, \citenamefont {Rottler}, \citenamefont {Durian}, \citenamefont {Kaxiras},\ and\ \citenamefont {Liu}}]{cubuk2015identifying}%
  \BibitemOpen
  \bibfield  {author} {\bibinfo {author} {\bibfnamefont {E.~D.}\ \bibnamefont {Cubuk}}, \bibinfo {author} {\bibfnamefont {S.~S.}\ \bibnamefont {Schoenholz}}, \bibinfo {author} {\bibfnamefont {J.~M.}\ \bibnamefont {Rieser}}, \bibinfo {author} {\bibfnamefont {B.~D.}\ \bibnamefont {Malone}}, \bibinfo {author} {\bibfnamefont {J.}~\bibnamefont {Rottler}}, \bibinfo {author} {\bibfnamefont {D.~J.}\ \bibnamefont {Durian}}, \bibinfo {author} {\bibfnamefont {E.}~\bibnamefont {Kaxiras}}, \ and\ \bibinfo {author} {\bibfnamefont {A.~J.}\ \bibnamefont {Liu}},\ }\href@noop {} {\bibfield  {journal} {\bibinfo  {journal} {Physical review letters}\ }\textbf {\bibinfo {volume} {114}},\ \bibinfo {pages} {108001} (\bibinfo {year} {2015})}\BibitemShut {NoStop}%
\bibitem [{\citenamefont {Gendelman}\ \emph {et~al.}(2015)\citenamefont {Gendelman}, \citenamefont {Jaiswal}, \citenamefont {Procaccia}, \citenamefont {Gupta},\ and\ \citenamefont {Zylberg}}]{gendelman2015shear}%
  \BibitemOpen
  \bibfield  {author} {\bibinfo {author} {\bibfnamefont {O.}~\bibnamefont {Gendelman}}, \bibinfo {author} {\bibfnamefont {P.~K.}\ \bibnamefont {Jaiswal}}, \bibinfo {author} {\bibfnamefont {I.}~\bibnamefont {Procaccia}}, \bibinfo {author} {\bibfnamefont {B.~S.}\ \bibnamefont {Gupta}}, \ and\ \bibinfo {author} {\bibfnamefont {J.}~\bibnamefont {Zylberg}},\ }\href@noop {} {\bibfield  {journal} {\bibinfo  {journal} {Europhysics Letters}\ }\textbf {\bibinfo {volume} {109}},\ \bibinfo {pages} {16002} (\bibinfo {year} {2015})}\BibitemShut {NoStop}%
\bibitem [{\citenamefont {Patinet}\ \emph {et~al.}(2016)\citenamefont {Patinet}, \citenamefont {Vandembroucq},\ and\ \citenamefont {Falk}}]{patinet2016connecting}%
  \BibitemOpen
  \bibfield  {author} {\bibinfo {author} {\bibfnamefont {S.}~\bibnamefont {Patinet}}, \bibinfo {author} {\bibfnamefont {D.}~\bibnamefont {Vandembroucq}}, \ and\ \bibinfo {author} {\bibfnamefont {M.~L.}\ \bibnamefont {Falk}},\ }\href@noop {} {\bibfield  {journal} {\bibinfo  {journal} {Physical review letters}\ }\textbf {\bibinfo {volume} {117}},\ \bibinfo {pages} {045501} (\bibinfo {year} {2016})}\BibitemShut {NoStop}%
\bibitem [{\citenamefont {Barbot}\ \emph {et~al.}(2018)\citenamefont {Barbot}, \citenamefont {Lerbinger}, \citenamefont {Hernandez-Garcia}, \citenamefont {Garc{\'\i}a-Garc{\'\i}a}, \citenamefont {Falk}, \citenamefont {Vandembroucq},\ and\ \citenamefont {Patinet}}]{barbot2018local}%
  \BibitemOpen
  \bibfield  {author} {\bibinfo {author} {\bibfnamefont {A.}~\bibnamefont {Barbot}}, \bibinfo {author} {\bibfnamefont {M.}~\bibnamefont {Lerbinger}}, \bibinfo {author} {\bibfnamefont {A.}~\bibnamefont {Hernandez-Garcia}}, \bibinfo {author} {\bibfnamefont {R.}~\bibnamefont {Garc{\'\i}a-Garc{\'\i}a}}, \bibinfo {author} {\bibfnamefont {M.~L.}\ \bibnamefont {Falk}}, \bibinfo {author} {\bibfnamefont {D.}~\bibnamefont {Vandembroucq}}, \ and\ \bibinfo {author} {\bibfnamefont {S.}~\bibnamefont {Patinet}},\ }\href@noop {} {\bibfield  {journal} {\bibinfo  {journal} {Physical Review E}\ }\textbf {\bibinfo {volume} {97}},\ \bibinfo {pages} {033001} (\bibinfo {year} {2018})}\BibitemShut {NoStop}%
\bibitem [{\citenamefont {Xu}\ \emph {et~al.}(2021)\citenamefont {Xu}, \citenamefont {Falk}, \citenamefont {Patinet},\ and\ \citenamefont {Guan}}]{xu2021atomic}%
  \BibitemOpen
  \bibfield  {author} {\bibinfo {author} {\bibfnamefont {B.}~\bibnamefont {Xu}}, \bibinfo {author} {\bibfnamefont {M.~L.}\ \bibnamefont {Falk}}, \bibinfo {author} {\bibfnamefont {S.}~\bibnamefont {Patinet}}, \ and\ \bibinfo {author} {\bibfnamefont {P.}~\bibnamefont {Guan}},\ }\href@noop {} {\bibfield  {journal} {\bibinfo  {journal} {Physical Review Materials}\ }\textbf {\bibinfo {volume} {5}},\ \bibinfo {pages} {025603} (\bibinfo {year} {2021})}\BibitemShut {NoStop}%
\bibitem [{\citenamefont {Adhikari}\ \emph {et~al.}(2023)\citenamefont {Adhikari}, \citenamefont {Sharma},\ and\ \citenamefont {Karmakar}}]{adhikari2023encoding}%
  \BibitemOpen
  \bibfield  {author} {\bibinfo {author} {\bibfnamefont {M.}~\bibnamefont {Adhikari}}, \bibinfo {author} {\bibfnamefont {R.}~\bibnamefont {Sharma}}, \ and\ \bibinfo {author} {\bibfnamefont {S.}~\bibnamefont {Karmakar}},\ }\href@noop {} {\bibfield  {journal} {\bibinfo  {journal} {arXiv preprint arXiv:2309.10682}\ } (\bibinfo {year} {2023})}\BibitemShut {NoStop}%
\bibitem [{\citenamefont {Scheibner}\ \emph {et~al.}(2020)\citenamefont {Scheibner}, \citenamefont {Souslov}, \citenamefont {Banerjee}, \citenamefont {Sur{\'o}wka}, \citenamefont {Irvine},\ and\ \citenamefont {Vitelli}}]{scheibner2020odd}%
  \BibitemOpen
  \bibfield  {author} {\bibinfo {author} {\bibfnamefont {C.}~\bibnamefont {Scheibner}}, \bibinfo {author} {\bibfnamefont {A.}~\bibnamefont {Souslov}}, \bibinfo {author} {\bibfnamefont {D.}~\bibnamefont {Banerjee}}, \bibinfo {author} {\bibfnamefont {P.}~\bibnamefont {Sur{\'o}wka}}, \bibinfo {author} {\bibfnamefont {W.~T.}\ \bibnamefont {Irvine}}, \ and\ \bibinfo {author} {\bibfnamefont {V.}~\bibnamefont {Vitelli}},\ }\href@noop {} {\bibfield  {journal} {\bibinfo  {journal} {Nature Physics}\ }\textbf {\bibinfo {volume} {16}},\ \bibinfo {pages} {475} (\bibinfo {year} {2020})}\BibitemShut {NoStop}%
\bibitem [{\citenamefont {Adhikari}\ and\ \citenamefont {Sastry}(2018)}]{adhikari2018memory}%
  \BibitemOpen
  \bibfield  {author} {\bibinfo {author} {\bibfnamefont {M.}~\bibnamefont {Adhikari}}\ and\ \bibinfo {author} {\bibfnamefont {S.}~\bibnamefont {Sastry}},\ }\href@noop {} {\bibfield  {journal} {\bibinfo  {journal} {The European Physical Journal E}\ }\textbf {\bibinfo {volume} {41}},\ \bibinfo {pages} {1} (\bibinfo {year} {2018})}\BibitemShut {NoStop}%
\bibitem [{\citenamefont {Keim}\ and\ \citenamefont {Medina}(2022)}]{keim2022mechanical}%
  \BibitemOpen
  \bibfield  {author} {\bibinfo {author} {\bibfnamefont {N.~C.}\ \bibnamefont {Keim}}\ and\ \bibinfo {author} {\bibfnamefont {D.}~\bibnamefont {Medina}},\ }\href@noop {} {\bibfield  {journal} {\bibinfo  {journal} {Science Advances}\ }\textbf {\bibinfo {volume} {8}},\ \bibinfo {pages} {eabo1614} (\bibinfo {year} {2022})}\BibitemShut {NoStop}%
\bibitem [{SM()}]{SM}%
  \BibitemOpen
  \href@noop {} {}\bibinfo {note} {See Supplemental Material at [URL] for details of the experimental setup and coupling between shear types, results of readout under different training conditions, and evidence of a box-shape memory effect}\BibitemShut {NoStop}%
\bibitem [{\citenamefont {Cuccia}\ \emph {et~al.}(2020)\citenamefont {Cuccia}, \citenamefont {Pothineni}, \citenamefont {Wu}, \citenamefont {M{\'e}ndez~Harper},\ and\ \citenamefont {Burton}}]{cuccia2020pore}%
  \BibitemOpen
  \bibfield  {author} {\bibinfo {author} {\bibfnamefont {N.~L.}\ \bibnamefont {Cuccia}}, \bibinfo {author} {\bibfnamefont {S.}~\bibnamefont {Pothineni}}, \bibinfo {author} {\bibfnamefont {B.}~\bibnamefont {Wu}}, \bibinfo {author} {\bibfnamefont {J.}~\bibnamefont {M{\'e}ndez~Harper}}, \ and\ \bibinfo {author} {\bibfnamefont {J.~C.}\ \bibnamefont {Burton}},\ }\href@noop {} {\bibfield  {journal} {\bibinfo  {journal} {Proceedings of the National Academy of Sciences}\ }\textbf {\bibinfo {volume} {117}},\ \bibinfo {pages} {11247} (\bibinfo {year} {2020})}\BibitemShut {NoStop}%
\bibitem [{\citenamefont {Lindeman}\ \emph {et~al.}(2023{\natexlab{a}})\citenamefont {Lindeman}, \citenamefont {Hagh}, \citenamefont {Ip},\ and\ \citenamefont {Nagel}}]{lindeman2023competition}%
  \BibitemOpen
  \bibfield  {author} {\bibinfo {author} {\bibfnamefont {C.~W.}\ \bibnamefont {Lindeman}}, \bibinfo {author} {\bibfnamefont {V.~F.}\ \bibnamefont {Hagh}}, \bibinfo {author} {\bibfnamefont {C.~I.}\ \bibnamefont {Ip}}, \ and\ \bibinfo {author} {\bibfnamefont {S.~R.}\ \bibnamefont {Nagel}},\ }\href@noop {} {\bibfield  {journal} {\bibinfo  {journal} {Physical review letters}\ }\textbf {\bibinfo {volume} {130}},\ \bibinfo {pages} {197201} (\bibinfo {year} {2023}{\natexlab{a}})}\BibitemShut {NoStop}%
\bibitem [{\citenamefont {Hagh}\ \emph {et~al.}(2022)\citenamefont {Hagh}, \citenamefont {Nagel}, \citenamefont {Liu}, \citenamefont {Manning},\ and\ \citenamefont {Corwin}}]{hagh2022transient}%
  \BibitemOpen
  \bibfield  {author} {\bibinfo {author} {\bibfnamefont {V.~F.}\ \bibnamefont {Hagh}}, \bibinfo {author} {\bibfnamefont {S.~R.}\ \bibnamefont {Nagel}}, \bibinfo {author} {\bibfnamefont {A.~J.}\ \bibnamefont {Liu}}, \bibinfo {author} {\bibfnamefont {M.~L.}\ \bibnamefont {Manning}}, \ and\ \bibinfo {author} {\bibfnamefont {E.~I.}\ \bibnamefont {Corwin}},\ }\href@noop {} {\bibfield  {journal} {\bibinfo  {journal} {Proceedings of the National Academy of Sciences}\ }\textbf {\bibinfo {volume} {119}},\ \bibinfo {pages} {e2117622119} (\bibinfo {year} {2022})}\BibitemShut {NoStop}%
\bibitem [{\citenamefont {Hexner}\ and\ \citenamefont {Levine}(2015)}]{hexner2015self}%
  \BibitemOpen
  \bibfield  {author} {\bibinfo {author} {\bibfnamefont {D.}~\bibnamefont {Hexner}}\ and\ \bibinfo {author} {\bibfnamefont {D.}~\bibnamefont {Levine}},\ }\href@noop {} {\bibfield  {journal} {\bibinfo  {journal} {Europhysics Letters}\ }\textbf {\bibinfo {volume} {109}},\ \bibinfo {pages} {30004} (\bibinfo {year} {2015})}\BibitemShut {NoStop}%
\bibitem [{\citenamefont {Lindeman}\ and\ \citenamefont {Nagel}(2021)}]{lindeman2021multiple}%
  \BibitemOpen
  \bibfield  {author} {\bibinfo {author} {\bibfnamefont {C.~W.}\ \bibnamefont {Lindeman}}\ and\ \bibinfo {author} {\bibfnamefont {S.~R.}\ \bibnamefont {Nagel}},\ }\href@noop {} {\bibfield  {journal} {\bibinfo  {journal} {Science Advances}\ }\textbf {\bibinfo {volume} {7}},\ \bibinfo {pages} {eabg7133} (\bibinfo {year} {2021})}\BibitemShut {NoStop}%
\bibitem [{\citenamefont {Lindeman}\ \emph {et~al.}(2023{\natexlab{b}})\citenamefont {Lindeman}, \citenamefont {Jalowiec},\ and\ \citenamefont {Keim}}]{lindeman2023isolating}%
  \BibitemOpen
  \bibfield  {author} {\bibinfo {author} {\bibfnamefont {C.~W.}\ \bibnamefont {Lindeman}}, \bibinfo {author} {\bibfnamefont {T.~R.}\ \bibnamefont {Jalowiec}}, \ and\ \bibinfo {author} {\bibfnamefont {N.~C.}\ \bibnamefont {Keim}},\ }\href@noop {} {\bibfield  {journal} {\bibinfo  {journal} {arXiv preprint arXiv:2306.07177}\ } (\bibinfo {year} {2023}{\natexlab{b}})}\BibitemShut {NoStop}%
\end{thebibliography}%





\include{SI}

\end{document}

\renewcommand{\thefigure}{S\arabic{figure}}
\setcounter{figure}{0} 
\renewcommand{\theequation}{S\arabic{equation}}
\setcounter{equation}{0}
\renewcommand{\thetable}{S\arabic{table}}
\setcounter{table}{0}

\title{Supplementary Material for \\ ``Multi-dimensional memory in sheared granular materials''}

\date{\today}

\maketitle

\renewcommand{\thefigure}{S\arabic{figure}}
\setcounter{figure}{0} 
\renewcommand{\theequation}{S\arabic{equation}}
\setcounter{equation}{0}

\section{Experimental details}

\begin{figure}
\centering
\includegraphics[width=15.5cm]{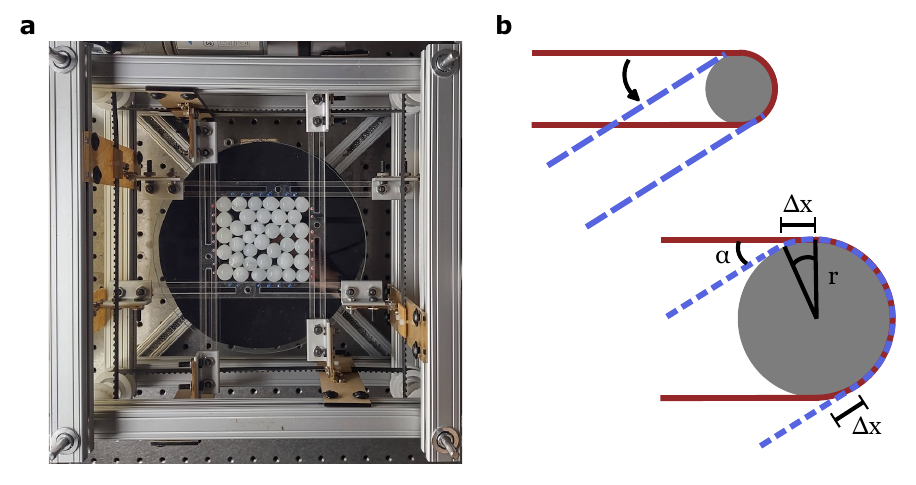}%
\caption{Experimental setup. (a) T-slot rails are used as outer frame, with one axle at each corner. Inner frame (laser-cut acrylic) is connected to belts via laser-cut sliders, which are constrained to slide along the t-slot rails and are driven by the belts. Particles are visible in the inner cell, and red and blue dots used to fit the boundaries are visible on the inner frame. Light source (not directly visible) is to the left. (b) Schematic zoom-ins of one pulley and belt to illustrate the coupling of $S2$ and $S1$ when $SB$ is applied.
}
\label{figsi:setup}
\end{figure}

Full setup used is shown in Fig.~\ref{figsi:setup}a. Particles sit on a glass plate and are confined by an acrylic inner frame, which is controlled to produce shear by the t-slot rail outer frame. Actuators drive shear by pushing and pulling a slider attached to one of the belts so that all four pulleys rotate ($SA$) and by pushing and pulling one axle along diagonal so that all four axles move in and out ($SB$).

\subsection{Coupling between $S2$ and $S1$ in application of $SB$}

When $S1$ is actuated via a linear actuator, it is intuitive that the walls will simply move together or apart with no change to the angles between them; $S2$ is unaffected. This is $SA$. In general one might also expect the reverse to be true, but because of the experimental setup driving $S2$ leads to some amount of $S1$ as well; this combination is defined as $SB$. To understand the origin of this coupling, consider the zoomed in schematic of one pulley and belt shown in Fig.~\ref{figsi:setup}b. Because the pulley is not free to rotate relative to the axle, the belt effectively rotates around the pulley, exposing new teeth on one side (in the schematic, this is the right side) and making new contact on the other (the left). The sliders that connect the inner frame to the belt are fixed in their connection to the belt so that they are forced to slide slightly as the belt changes its configuration. 

With no hysteresis, this slide distance $\Delta x$ can be calculated as the length of newly exposed belt (or, equivalently, belt newly in contact with the pulley), a function of the change in angle of the wall $\alpha$ (between $2.5^{\circ}$ and $3^{\circ}$ for a typical training cycle) and the radius $r$ of the pulley (about $1$ in). This gives a sliding distance of 
$$\Delta x = \frac{3^{\circ}}{360^{\circ}} \times 2 \pi \times 1 \hspace{1 mm} \textrm{in} \sim 0.05 \hspace{1 mm} \textrm{in}.$$
Comparing this with the typical sliding distance for one training cycle $\Delta x_{train} \sim 0.09 \hspace{1 mm} \textrm{in}$, we find that the magnitude of $S1$ when $SB$ is applied should be about half that when $SA$ is applied. In practice the ratio is closer to 0.3, likely due to friction preventing the sliders from moving freely.

\subsection{Strain magnitudes}

For two-dimesional measurements (that is, the results presented in Fig.~5), the $SA$ and $SB$ training amplitudes were both $5\%$.

For one-dimensional measurements, $SA$ training magnitudes are $7\%$ (or $9\%$ for the ``larger" amplitude scenario) and $SB$ training magnitudes are $5\%$ (or $7\%$ for the ``larger" amplitude scenario).  

\section{Readout under other conditions}

\begin{figure}
\centering
\includegraphics[width=17.2cm]{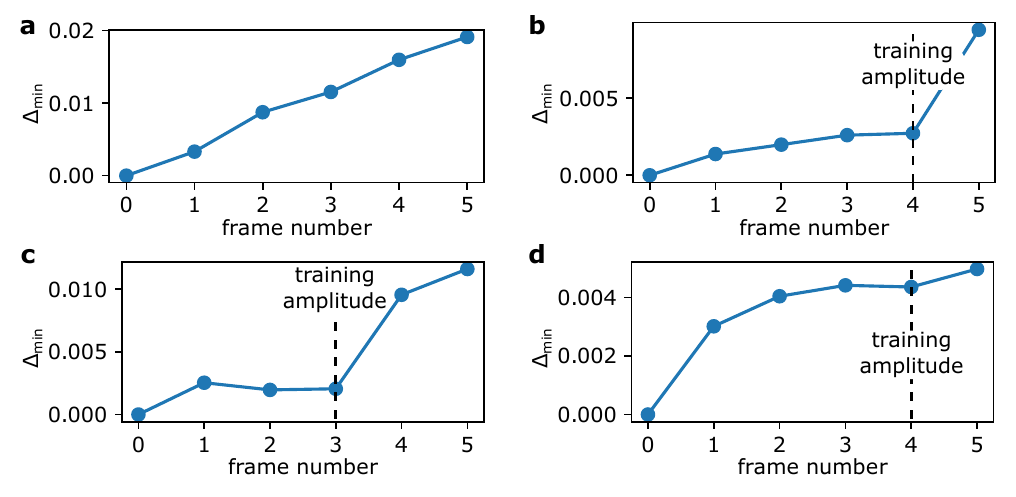}%
\caption{(a) Readout via $SB$ of untrained packings and (b) readout via $SB$ of packings trained with $S2$ at larger amplitude than shown in the main text. (c,d) Readout via $SA$ for packings trained with $SA$ at two different amplitudes.
All sets of data are an average of five experiments under identical conditions.
}
\label{figsi:other-training}
\end{figure}

Figure~\ref{figsi:other-training}a shows readout for packings that have not been trained at all. Readout is monotonically increasing, revealing the lack of training. We also demonstrate the ability to train values other than that shown in the main text. Figure~\ref{figsi:other-training}b shows readout after training at a larger amplitude than the examples shown in the main text; cusp appears when the readout amplitude reaches that larger value. Similar training is also possible with $S1$, as shown in Fig.~\ref{figsi:other-training}(c,d).

\section{Box memory effects}

%In an ideal experiment, the box would return back to its exact initial shape after each cycle of shear. However, in practice, even the ``stroboscopic'' box shape is not always identical. We find that this stroboscopic box shape exhibits some memory of its own during readout. We can quantify the box shape via $S1$ and $S2$, which in general are much smaller than the applied values during training (typically less than $0.5\%$ in stroboscopic measurements compared with $\sim 5 \%$ at the maxima of training cycles). Figure~\ref{figsi:box-shape} shows $S1$ and $S2$ during readout with (a) all points plotted so that $S2$ exhibits the characteristic ``ring-up'' protocol show in Fig.~2b of the main text and (b) with only the stroboscopic points shown, so that small changes in the box shape are highlighted. In (c-f), the absolute value of the change in $S1$ is shown, revealing that in many cases a minimum occurs where when the system returns to the training amplitude: the box comes closest back to its trained shape at the readout cycle with the same amplitude as training. 

\begin{figure}
\centering
\includegraphics[width=17.2cm]{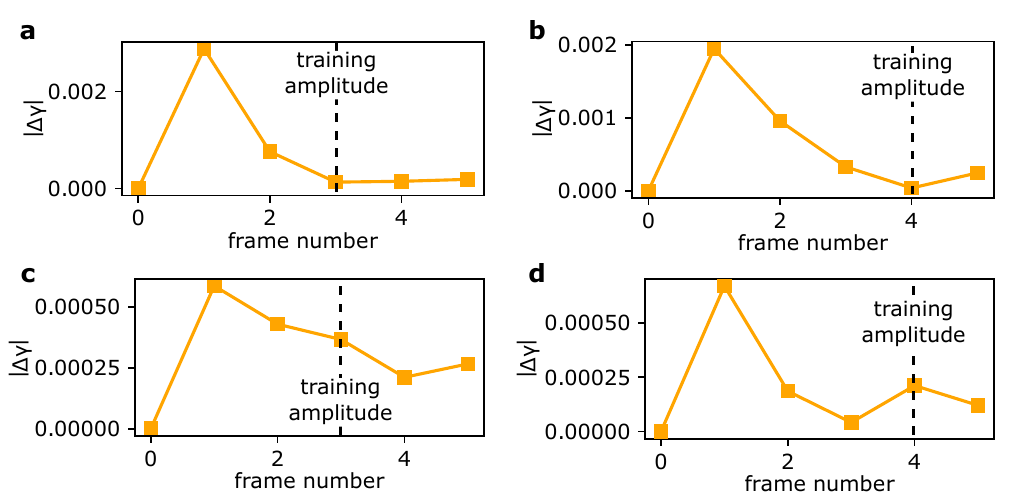}%
\caption{Stroboscopic measured change in $S1$ strain values (a) for the experiments reported as purple squares in Fig.~4b (trained and read out with loops), (b) for the experiments reported in Fig.~\ref{figsi:other-training}b (trained and read out with $SB$), and (c,d) for the experiments reported in Fig.~\ref{figsi:other-training}(c,d) (trained and read out with $SA$).
}
\label{figsi:box-shape}
\end{figure}

As discussed in the main text, the box exhibits some memory of its own. Figure~\ref{figsi:box-shape} shows the stroboscopic $| \Delta \gamma |$ values introduced in Fig.~5 of the main text for several other sets of experiments. All cases shown in Fig.~\ref{figsi:box-shape} are cases where a memory was present in the readout signal; about half exhibit box memory.

%\bibliography{apssamp}